\documentclass[aps, prb, twocolumn, superscriptaddress, showpacs]{revtex4}
\usepackage{amssymb}
\usepackage{amsmath}
\usepackage{bm}
\usepackage[dvips]{graphicx}
\usepackage{dcolumn}
\usepackage{longtable}
\usepackage{color}

\begin{document}

\title{Universality and Critical Behavior at the Critical-End-Point on Itinerant-Metamagnet UCoAl}
\author{K.~Karube}
\altaffiliation{karube@scphys.kyoto-u.ac.jp}
\author{T.~Hattori}
\author{S.~Kitagawa}
\author{K.~Ishida}
\affiliation{Department of Physics, Graduate School of Science, Kyoto University, Kyoto 606-8502, Japan}
\author{N.~Kimura}
\author{T.~Komatsubara}
\affiliation{Department of Physics, Graduate School of Science, Tohoku University, 
Sendai, 980-8578, Japan,}
\affiliation{Center for Low Temperature Science, Tohoku University, Sendai,980-8578, Japan.}
\date{\today}

\begin{abstract}
We performed nuclear-magnetic-resonance (NMR) measurements on itinerant-electron metamagnet UCoAl in order to investigate the critical behavior of the magnetism near a metamagnetic (MM) critical endpoint (CEP). 
We derived $c$-axis magnetization $M_c$ and its fluctuation $S_c$ from the measurements of Knight shift and nuclear spin-lattice relaxation rate $1/T_1$ as a function of the $c$-axis external field ($H_c$) and temperature ($T$). 
We developed contour plots of $M_c$ and $S_c$ on the $H_c$ - $T$ phase diagram, and observed the strong divergence of $S_c$ at the CEP. 
The critical exponents of $M_c$ and $S_c$ near the CEP are estimated, and found to be close to the universal properties of a three-dimensional (3-D) Ising model. We indicate that the critical phenomena at the itinerant-electron MM CEP in UCoAl have a common feature as a gas-liquid transition.
\end{abstract}

\pacs{74}
\maketitle
\section{Introduction}
Magnetic properties on Uranium (U) compounds have attracted much interest since novel phenomena such as a hidden order in URu$_2$Si$_2$\cite{PalstraPRL,MaplePRL} and superconducting ferromagnet in UGe$_2$\cite{SaxenaNature}, URhGe\cite{AokiNature} and UCoGe\cite{HuyPRL} were reported.
In this paper, we report magnetic properties on UCoAl possessing the hexagonal ZrNiAl-type structure shown in Fig. 1. 
UCoAl shows a characteristic first-order metamagnetic (MM) transition at low temperatures\cite{Andreev,Mushnikov}. 
The ground state of UCoAl is paramagnetic (PM) with a strong Ising-like anisotropy (the easy axis is the $c$-axis), and magnetization ($M$) along the $c$-axis $M_c$ shows an abrupt jump with hysteresis below 10 K, when relatively small external magnetic fields between 0.7 - 1 T are applied along the $c$-axis\cite{Andreev,Mushnikov}. 
This is the first-order MM transition from the PM state to the ferromagnetic (FM) state, but it is noted that UCoAl is an itinerant-electron metamagnet originating from U-5$f$ electrons. 
The induced FM moments ($\sim$ 0.3 $\mu_\mathrm{B}$) are much smaller than the effective moments ($\sim$ 1.8 $\mu_\mathrm{B}$) evaluated from the Curie-Weiss behavior above 40 K, and $M_c$ becomes larger with applied fields even above the MM transition. 
This first-order MM transition in UCoAl terminates at a finite temperature critical-end-point (CEP), ($\mu_{0}H_c$, $T$)$_{\rm CEP} \sim$ (1 T, 12 K)\cite{Mushnikov,Nohara,Aoki}, as shown schematically in Fig. 2 (a). 
It was suggested that UCoAl at ambient pressure is a similar state as UGe$_2$ at $P$ $\sim$ 2 GPa, since the presence of the similar CEP was reported on UGe$_2$ in the pressure region of 1.5 $<$ $P$ $<$ 3 GPa.\cite{Taufour}
Above the CEP, the borderline of the first-order transition becomes bleared and the PM state continuously connects to the FM state as crossover. 
This CEP reminds us of a gas-liquid transition (see in Fig.~2 (b)), where the order parameter and the tuning parameter are density of molecules and pressure, respectively\cite{Kadanoff}. 
Therefore, an important and fundamental question is that what kind of universality class is observed near the itinerant-electron MM CEP, since its universality has never been reported so far and the change of the electronic structure was recently suggested in the field-induced FM state\cite{Matsuda,Yamaji}.
%%%%%%%%%%%%%%%%%%  FIG1   %%%%%%%%%%%%%%
\begin{figure}[tbp]
\begin{center}
\includegraphics[width=7cm]{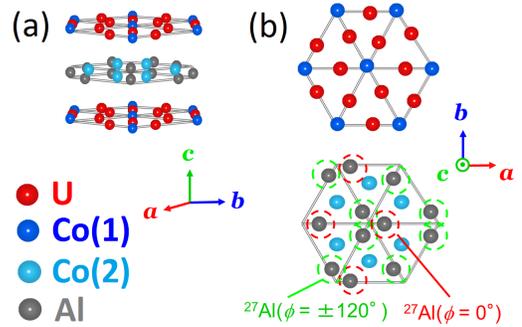}
\end{center}
\caption{(Color online) (a) Hexagonal crystal structure of UCoAl composed by U-Co(1) layer and Co(2)-Al layer alternatively stacking along $c$-axis. (b) U-Co(1) layer and Co(2)-Al layer from the view of $c$-axis. When the external field is applied along $a$-axis, there exist two inequivalent Al site, marked by red circle for $^{27}$Al($\phi$ = 0$^\circ$) and light green circle for $^{27}$Al($\phi$ = $\pm$120$^\circ$), where $\phi$ is the angle between the direction of external field and EFG second principal axis in the basal plane.}
\label{fig1}
\end{figure}
%%%%%%%%%%%%%%%%%%%  FIG1   %%%%%%%%%%%%%

%%%%%%%%%%%%%%%%%%%  FIG2   %%%%%%%%%%%%%
\begin{figure}[t]
\begin{center}
\includegraphics[width=8cm]{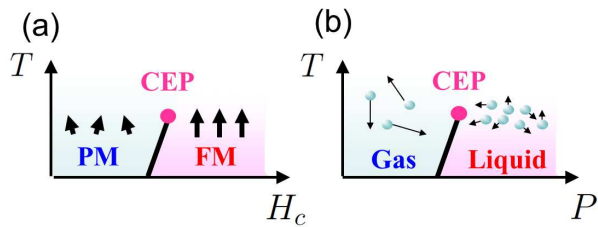}
\end{center}
\caption{(Color online) Schematic figures of (a) a metamagnetic transition and (b) a gas-liquid transition. Both figures have a critical-end-point (CEP) at a finite temperature. Below the CEP, two phases are separated by the first-order transition line.  
}
\label{fig2}
\end{figure}
%%%%%%%%%%%%%%%%%%%  FIG2   %%%%%%%%%%%%%

We point out that a precise field-tuned NMR study on the itinerant metamagnet UCoAl is an ideal experiment for investigating the physical properties around the CEP, since (I) the MM transition in UCoAl occurs at relatively smaller magnetic fields, (II) magnetic fields (tuning parameter) can be controlled continuously and precisely, and (III) magnetization (order parameter) and its dynamical fluctuations can be detected by NMR measurements microscopically. 
For the study of a first-order transition, microscopic measurements are crucial since they can discriminate between homogeneous and inhomogeneous (coexisting) states explicitly. 

\section{Experiment}

We performed $^{27}$Al-NMR measurements on a single-crystal UCoAl. 
A single-crystal UCoAl sample was synthesized by Czochralski-pulling method in a tetra-arc furnace, and was cut as rectangular cubic shape with 1.5($a$-axis) $\times$ 3.2($b$-axis) $\times$ 1.7($c$-axis) mm$^3$. 
This single-crystal UCoAl was used for angle-resolved $^{27}$Al-NMR measurement in order to investigate $H_c$ (magnetic field along the $c$-axis) and temperature dependences of Knight shift $K$ and nuclear spin-lattice relaxation rate $1/T_1$, by controlling the angle $\theta$ between the $c$-axis and the external field $H$ in the $ac$-plane.
NMR measurements were carried out at two frequencies of 29.8 MHz ($\mu_0 H$($^{27} K$ = 0) = 2.686 T) and 49.1 MHz ($\mu_0 H$($^{27} K$ = 0) = 4.426 T). 
The typical $H$ swept $^{27}$Al and $^{59}$Co-NMR spectra obtained under $H$ parallel to the $a$-axis ($\theta = 90^{\circ}$) at 29.8 MHz is shown in Fig. 3.
All NMR peaks are well identified as shown in Fig. 3.
In the field along the $a$-axis, there exist two inequivalent $^{27}$Al sites, denoted as $^{27}$Al($\phi$ = 0 $^\circ$) and $^{27}$Al($\phi$ = $\pm$120 $^\circ$), where $\phi$ is the angle between the direction of external field and electric field gradient (EFG) second principal axis in basal plane. These two inequivalent $^{27}$Al nuclei ($I$ = 5/2) each provide four quadrupole satellites at different resonance field as calculated by the following first-order perturbation formula for $m\leftrightarrow(m-1)$ transition,
\begin{eqnarray}
\lefteqn{\Delta\nu_{m \leftrightarrow m-1}} \nonumber \\
&=&\frac{\nu_{zz}}{2} \left(m-\frac{1}{2} \right) \left\{ ( 3{\cos}^2 \theta -1)-\eta\:{\sin}^2 \theta \cos 2\phi \right\}
\end{eqnarray}
where $\nu_{zz}$ is a quadrupole resonance frequency along the EFG principle axis ($c$-axis), and $\eta$, defined as $|\nu_{xx}-\nu_{yy}|/\nu_{zz}$, is an asymmetry parameter about the EFG principal axis.
From the observed $^{27}$Al-NMR spectra and above theoretical equation we obtained the quadrupole parameters for the $^{27}$Al nucleus as shown in Table I.
The quadrupole parameters of two Co sites in UCoAl are also listed in Table I \cite{Iwamoto}.

$K$ and $1/T_1$ were measured at a central peak of $^{27}$Al-NMR spectra, corresponding to the transition between the nuclear spin states $I$ = 1/2 and -1/2. 
For the measurements of $1/T_1$, nuclear magnetization after saturation pulses can be fitted consistently with the theoretical function in whole measurements. 
The angle $\theta$ was controlled by using a split-coil superconducting magnet and a rotator with the precision of 0.5$^\circ$ .

\begin{table}[h]
\caption{NQR parameters of $^{59}$Co(1), $^{59}$Co(2) and $^{27}$Al nuclei in UCoAl}
\begin{center}
\begin{tabular}{cccc} \hline
Nucleus & $\nu_{zz}$ (MHz) & $\eta$ & Reference \\ \hline
$^{59}$Co(1) & 0.695 & 0 & Iwamoto \textit{et al}\cite{Iwamoto}\\ 
$^{59}$Co(2) & 4.32 & 0 & Iwamoto \textit{et al}\cite{Iwamoto} \\ 
$^{27}$Al & 0.385 & 0.327 & This work\\\hline
\end{tabular}
\end{center}
\end{table}

%%%%%%%%%%%%%%%%%%%  FIG3   %%%%%%%%%%%%%%%%
\begin{figure}[t]
\begin{center}
\includegraphics[width=9cm]{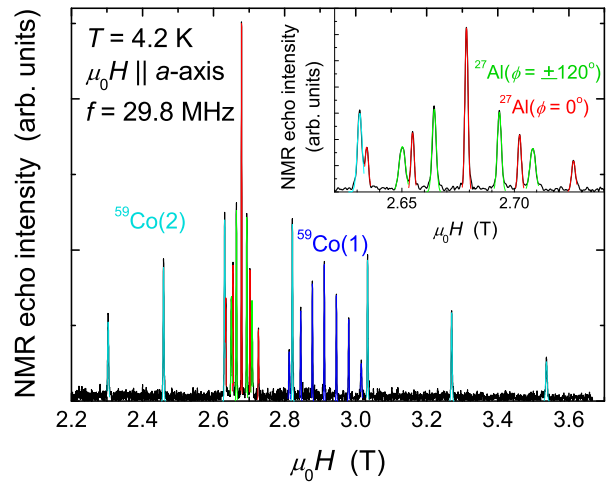}
\end{center}
\caption{(Color online) Field-swept NMR spectra at $T$ = 4.2 K in the field applied along the $a$-axis. The $^{59}$Co-NMR spectra are shown with dark and light blue for $^{59}$Co(1) and $^{59}$Co(2) sites, and $^{27}$Al-NMR spectra, which split into two sites in the $H$ parallel to the $a$-axis, are shown with red and green for $^{27}$Al($\phi$ = 0$^\circ$) and $^{27}$Al($\phi$ = $\pm$120$^\circ$), respectively.  Here, $\phi$ is the angle between the direction of external field and EFG second principal axis in basal plane. 
}
\label{fig3}
\end{figure}
%%%%%%%%%%%%%%%%%%%  FIG3   %%%%%%%%%%%%%%%%%%%%%%%

%%%%%%%%%%%%%%%%%%%  FIG4   %%%%%%%%%%%%%%%%%%%%%%%%%%%%%%%%%%%
\begin{figure}[t]
\begin{center}
\includegraphics[width=9cm]{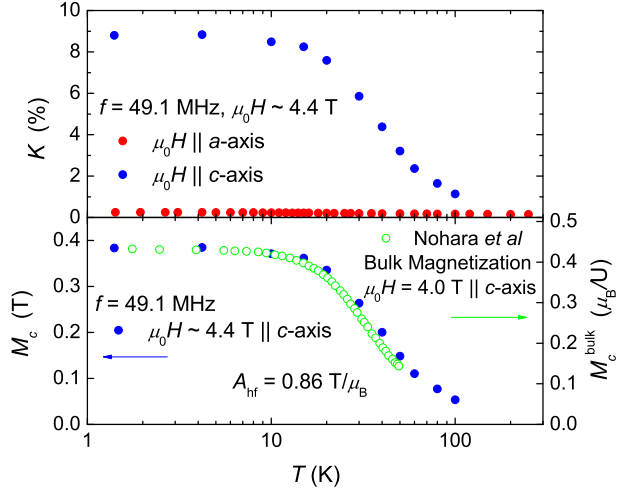}
\end{center}
\caption{(Color online) (Upper panel) Temperature dependence of Knight-shift $K$ in the field $\mu_0H \sim$ 4.4 T applied along the $a$-axis (red circle) and the $c$-axis (blue circle). (Lower panel) Temperature dependence of magnetization along the $c$-axis $M_c$ evaluated with the above $^{27}$Al-NMR Knight-shift results (blue circle) and bulk magnetization along the $c$-axis ($M_c^{\rm bulk}$) in $\mu_0H_c$ = 4.0 T (green open circle)\cite{Nohara}. $M_c$ and $M_c^{\rm bulk}$ were well scaled with the hyperfine-coupling constant $A_\mathrm{hf} = 0.86$ T/$\mu_\mathrm{B}$.
}
\label{fig4}
\end{figure}
%%%%%%%%%%%%%%%%%%%%  FIG4   %%%%%%%%%%%%%%%%%%%%%%%%%%%%%%%%%%%

%%%%%%%%%%%%%%%%%%%%  FIG5   %%%%%%%%%%%%%%%%%%%%%%%%%%%%%%%%%%%
\begin{figure}[t]
\begin{center}
\includegraphics[width=9cm]{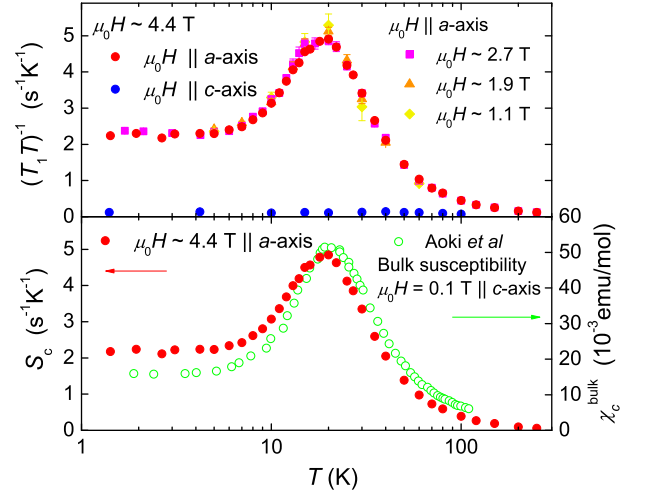}
\end{center}
\caption{(Color online) (Upper panel) Temperature dependence of $(T_1T)^{-1}$ in the field $\mu_0H \sim$ 4.4 T applied along the $a$-axis (red circle) and $c$-axis (blue circle). $(T_1T)^{-1}$ in the different fields along the $a$-axis are also plotted. (Lower panel) Temperature dependence of magnetic fluctuation $S_c$ along the $c$-axis evaluated with above $(T_1T)^{-1}$ results, together with bulk magnetic susceptibility $\chi_c^{\rm bulk}$ in $\mu_0H \sim 0.1$ T along the $c$-axis (green open symbol)\cite{Aoki}. $S_c$ and $\chi_c^{\rm bulk}$ are well scaled with each other.
}
\label{fig5}
\end{figure}
%%%%%%%%%%%%%%%%%%%  FIG5   %%%%%%%%%%%%%%%%%%%%%%%%%%%%%%%%%%%

%%%%%%%%%%%%%%%%%%%%%%  FIG6   %%%%%%%%%%%%%%%%%%%%%%%%%%%%%%%%%%%
\begin{figure}[t]
\begin{center}
\includegraphics[width=7cm]{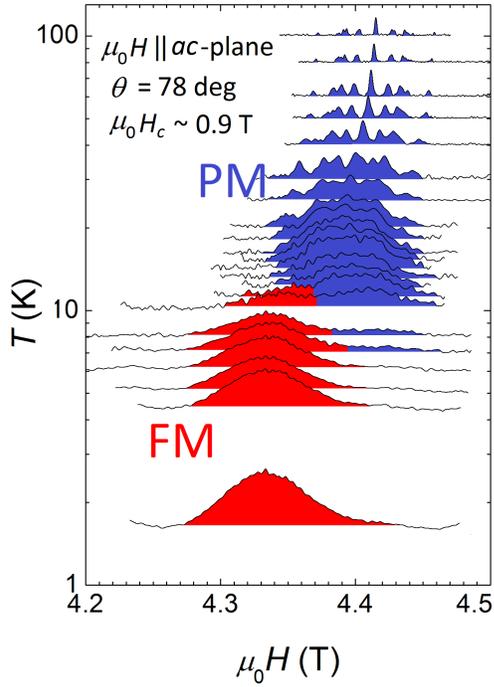}
\end{center}
\caption{(Color online) Temperature scanned $^{27}$Al-NMR spectra at the angle $\theta$ = 78$^\circ$ ($\mu_0 H_c$ $\sim$ 0.9 T) in the first-order transition region. $^{27}$Al-NMR spectra colored by blue and red represents PM and FM components, respectively. Between 11 K and 7 K both of PM and FM signals appear, where PM and FM components coexist.
}
\label{fig6}
\end{figure}
%%%%%%%%%%%%%%%%%%%%%%  FIG6   %%%%%%%%%%%%%%%%%%%%%%%%%%%%%%%%%%%

%%%%%%%%%%%%%%%%%%%%%%  FIG7   %%%%%%%%%%%%%%%%%%%%%%%%%%%%%%%%%%%
\begin{figure}[t]
\begin{center}
\includegraphics[width=7cm]{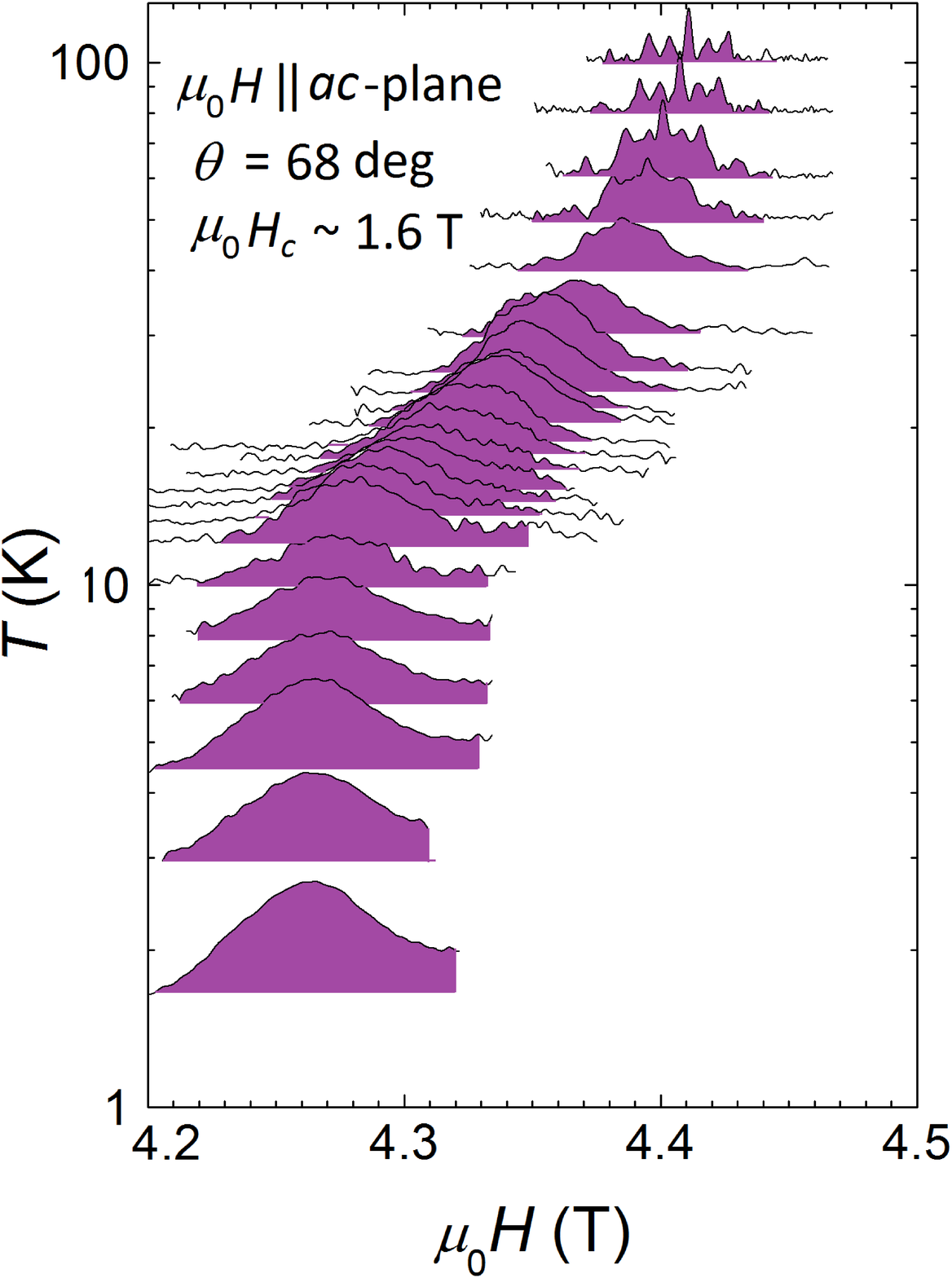}
\end{center}
\caption{(Color online) Temperature scanned $^{27}$Al-NMR spectra at the angle $\theta$ = 68$^\circ$ ($\mu_0 H_c$ $\sim$ 1.6 T) in the crossover region. $^{27}$Al-NMR spectra colored by purple continuously shift.
}
\label{fig7}
\end{figure}
%%%%%%%%%%%%%%%%%%%%%%  FIG7   %%%%%%%%%%%%%%%%%%%%%%%%%%%%%%%%%%%

\section{Experimental Results and Discussion}
\subsection{Ising-type anisotropy in magnetization and ferromagnetic fluctuations}
NMR Knight shift, which is proportional to microscopic spin susceptibility at the nuclear site, and nuclear spin-lattice relaxation rate $1/T_1$, probing the electronic spin dynamics, are measured down to 1.5 K and in the magnetic field ($H$) up to 4.4 T. 
The upper panel of Fig. 4 shows $T$ dependences of Knight shift in $\mu_0 H \sim 4.4$ T along the $a$- and $c$-axes. 
The Ising-type strong anisotropy, $K_a \ll K_c$ was observed. 
The Knight-shift $K(\theta)$ at the angle $\theta$ between an external field $H$ and the $c$-axis is expressed as the relation of $K(\theta) = K_c \cos^2{\theta} + K_a \sin^2{\theta}$. 
Using this relation, the $c$-axis magnetization at $^{27}$Al nucleus site was evaluated as
\begin{eqnarray}
M_c^\prime = K_c \mu_0H_c = \mu_0H_c [K(\theta) - K_a \sin^2{\theta}] \cos^{-2}{\theta}.
\end{eqnarray}
Here we labeled $M_c^\prime$ instead of $M_c$ to make clear that this magnetization calculated from Knight shift contains demagnetization effect.
Now we need to estimate the $M_c$ without demagnetization effect.
The shape of the sample in the applied field is the almost same rectangle ($ac$-plane vs $b$-axis, the ratio of the length is $b/a(c)$ $\sim$ 2).
Thus, the demagnetization field in the unit of $\mu_\mathrm{B}$ is estimated as $D$ = 0.064 T/$\mu_\mathrm{B}$. 
Using the hyperfine-coupling constant $A_\mathrm{hf}$ between $^{27}$Al nucleus and U-5$f$ electron, the demagnetization factor $D$ and bulk magnetization $M_c^\mathrm{bulk}$ from U-5$f$ moment along the $c$-axis without the demagnetization effect, $M_c^\prime$ is written as follows,
\begin{eqnarray}
M_c^\prime= A_\mathrm{hf} M_c^\mathrm{bulk} - D M_c^\mathrm{bulk} \equiv A_\mathrm{hf}^\prime M_c^\mathrm{bulk}
\end{eqnarray}
Here, we define the hyperfine-coupling constant with demagnetization as $A_\mathrm{hf}^\prime \equiv A_\mathrm{hf} - D$.
$A_{\rm hf}^\prime $ is estimated as 0.80 T/$\mu_B$ from $K_c \mu_0H_c$ vs $M_c^{\rm bulk}$ plot. 
Thus, $M_c$ without demagnetization is corrected as follows,
\begin{eqnarray}
M_c = A_\mathrm{hf} M_c^\mathrm{bulk}=(1+D/A_\mathrm{hf}^\prime)M_c^\prime \sim 1.08  M_c^\prime
\end{eqnarray}
The lower panel of Fig. 4 shows the $T$ dependence of the calculated $M_c$, in which $M_c^{\rm bulk}$ measured in the field $\mu_0 H = 4.0$ T is also plotted\cite{Nohara}. 
The calculated $M_c$ is well scaled to $M_c^{\rm bulk}$ with the hyperfine-coupling constant $A_{\rm hf}$= 0.86 T/$\mu_B$. 
This verifies that $^{27}$Al-NMR results are determined by the U - 5$f$ magnetic properties. 

The upper panel of Fig. 5 shows the $T$ dependence of $(T_1T)^{-1}$ in the field ($\mu_0~H \sim 4.4$ T) along the $a$- and $c$-axes, respectively. 
In contrast with the Knight-shift behavior, $(T_1T)^{-1}$ along the $a$-axis [$(T_1T)^{-1}_a$] shows temperature dependence with a broad maximum around 20 K and the Korringa relation ($T_1T$ = const.) below 7 K, but $(T_1T)^{-1}$ along the $c$-axis [$(T_1T)^{-1}_c$] is nearly constant with a small value. 
This is because $(T_1T)^{-1}$ probes hyperfine-field fluctuations perpendicular to the applied fields. 
Thus, $(T_1T)^{-1}$ measured in a field along the $i$ direction is expressed as $(T_1T)^{-1}_i \equiv S_j + S_k$, where $S_{j, (k)}$ is magnetic fluctuations along the $j, (k)$ direction $\left[ S_{j, (k)} \propto \sum_q |S_{j, (k)}(q,\omega \sim 0)| \right] $ 
and $i, j$ and $k$ directions are mutually orthogonal.  
In addition, $(T_1T)^{-1}$ at the angle $\theta$ is expressed as, $(T_1T)^{-1}(\theta) = (T_1T)^{-1}_c \cos^2{\theta} + (T_1T)^{-1}_a \sin^2{\theta}$. 
If we assume that the in-plane magnetic fluctuations are isotropic [$(T_1T)^{-1}_c \equiv S_a + S_b \sim 2S_a$], $S_c$ is evaluated as, 
\begin{eqnarray}
S_c = \left[(T_1T)^{-1}(\theta) - \frac{(1+\cos^2{\theta}) (T_1T)^{-1}_c}{2}\right] \sin^{-2}{\theta}
\label{T1Teq}
\end{eqnarray}
from the measurements of $(T_1T)^{-1}_c$ and $(T_1T)^{-1}(\theta)$. 
The lower panel of Fig.~5 shows the $T$ dependence of $S_c$ calculated with eq.(\ref{T1Teq}) by using the upper-panel $(T_1T)^{-1}$ data. 
In the figure, the $T$ dependence of bulk magnetic susceptibility $\chi_c^{\rm bulk}$ measured in the field $\mu_0H = 0.1$ T along the $c$-axis was also plotted\cite{Aoki}. 
The good scaling between $S_c$ and $\chi_c^{\rm bulk}$ indicates that the magnetic fluctuations along the $c$-axis is completely insensitive to the field along the $a$-axis and suggests that the magnetic fluctuations originating from the U - 5$f$ electrons possess the three dimensional (3-D) FM fluctuations, since the relation of ($(T_1T)^{-1} \propto \chi$) was anticipated in the self-consistent-renormalization (SCR) theory when 3-D FM fluctuations are dominant\cite{Moriya}. 
The presence of the 3-D FM fluctuations is also consistent with the resistivity data [$\rho(T) \propto T^{5/3}$] measured in zero field\cite{Havela}. 
We comment on that the similar Ising magnetic fluctuations were observed in a ferromagnetic superconductor UCoGe\cite{Ihara,Hattori}.  

%%%%%%%%%%%%%%%%%%%%%%%  FIG8   %%%%%%%%%%%%%%%%%%%%%%%%%%%%%%%%%%%
\begin{figure}[b]
\begin{center}
\includegraphics[width=9cm]{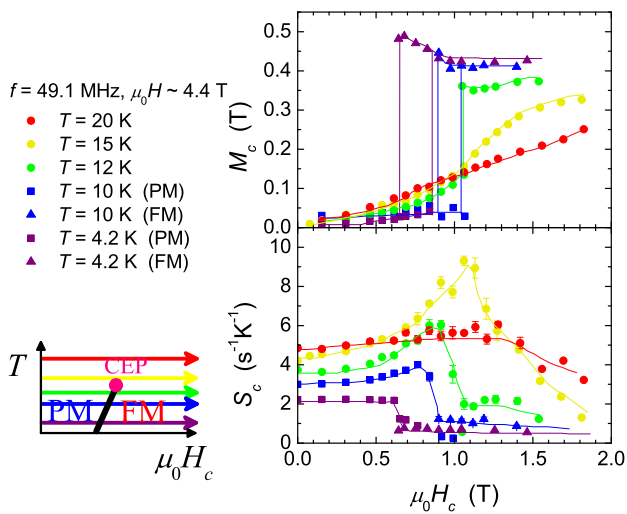}
\end{center}
\caption{(Color online) $H_c$ dependence of $M_c$ and $S_c$ at several fixed temperatures. At 4.2 K and 10 K, PM and FM components separated by the first-order transition are denoted as square and triangle symbols, respectively. At the other temperatures, data points are denoted as circle symbols. The measurement scans are shown in the schematic $H_c$ - $T$ phase diagram by colored arrows. Each color of an arrow corresponds to that of data points.
}
\label{fig8}
\end{figure}
%%%%%%%%%%%%%%%%%%%%%%%  FIG8   %%%%%%%%%%%%%%%%%%%%%%%%%%%%%%%%%%%

\subsection{$H_c$ dependence of $M_c$ and $S_c$}
In order to investigate the dependence of $M_c$ and $S_c$ against magnetic fields along the $c$-axis ($H_c$), we measured $K(\theta)$ and $(T_1T)^{-1}(\theta)$ by controlling the angle $\theta$ in the $ac$ plane [the $c$-axis ($\theta = 0^{\circ}$) and  the $a$-axis ($\theta = 90^{\circ}$)]. 
This is because that applied magnetic field is decomposed to the fields along the $a$-axis ($H_a$)  and $c$-axis ($H_c$) with respect to the sample, and $M_c$ and $S_c$ are not affected by $H_a$, but very sensitive to $H_c$.   
This experimental condition enabled us to control $H_c (= H\cos{\theta})$ continuously with a fixed NMR frequency $f_0 = 49.1$ MHz, and thus to scan $H_c$ across the CEP.
Figures 6 and 7 show temperature variation of the $^{27}$Al-NMR spectra below and above the critical field of $\mu_0H_{c} \sim$ 1 T, respectively. 
NMR spectra obtained below $\mu_0H_{c} \sim$ 1 T (Fig.~6) show a discontinuous shift around 10 K with a coexistence of the PM and FM spectra. The field difference between the PM and FM signals is much larger than the demagnetization field ($-DM_c^\mathrm{bulk}$ $\sim$ $-$0.02 T), and the demagnetization field works to reduce the field difference between two signals. 
Therefore, we consider that the coexistence of the PM and FM signals is not due to the demagnetization effect, but due to the first-order transition.
On the other hand, NMR spectra obtained above $\mu_0H_{c} \sim$ 1 T (Fig.~7) shows a continuous shift with decreasing temperature.
The NMR measurement is a powerful technique to distinguish between the first-order transition and crossover in metamagnetic behavior. 

%%%%%%%%%%%%%%%%%%%%%%%  FIG9   %%%%%%%%%%%%%%%%%%%%%%%%%%%%%%%%%%%
\begin{figure}[b]
\begin{center}
\includegraphics[width=9.3cm]{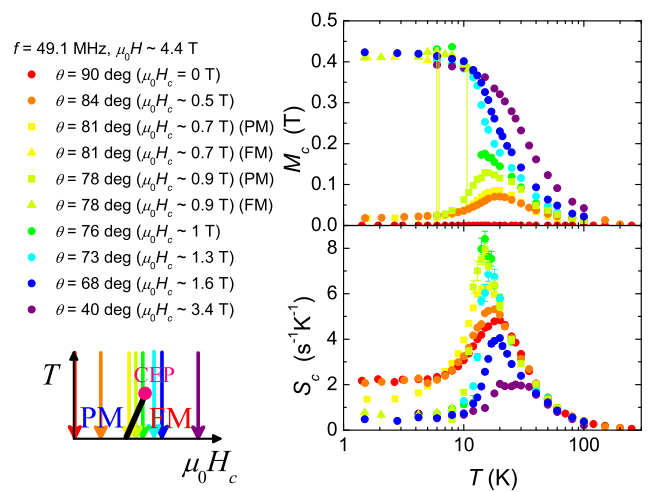}
\end{center}
\caption{(Color online) Temperature dependence of $M_c$ and $S_c$ for several fixed $\theta$ (i.e. $H_c$). At $\mu_0H_c$ = 0.7 T and 0.9 T, the PM and FM components separated by the first-order transition are denoted as square and triangle symbols, respectively. At other $H_c$, data points are denoted as circle symbols. The measurement scans are shown in the schematic $H_c$ - $T$ phase diagram by colored arrows. Each color of an arrow corresponds to that of data points.
}
\label{fig9}
\end{figure}
%%%%%%%%%%%%%%%%%%%%%%%  FIG9   %%%%%%%%%%%%%%%%%%%%%%%%%%%%%%%%%%%
Figure 8 shows the $H_c$ dependence of $M_c$ (upper panel) and $S_c$ (lower panel) at several fixed temperatures. 
At $T$ = 4.2 and 10 K, $M_c$ shows a first-order transition from the PM to FM state with a coexisting region where NMR signals from the PM and FM states were observed. 
Above $T$ = 12 K, the first-order transition disappears and $M_c$ continuously changes against $H_c$. 
Correspondingly, $S_c$ at 4.2 and 10 K suddenly drops at the MM transition field without a notable divergence, but $S_c$ of the PM component also drops to the same value as that of the FM component. 
At 12 and 15 K, very close to a critical temperature, $S_c$ exhibits a pronounced peak around $\mu_0H_c \sim 1$ T, and the peak becomes suppressed by getting away from the critical temperature. 

%%%%%%%%%%%%%%%%%%%  FIG10   %%%%%%%%%%%%%%%%%%%%%%%%%%%%%%%%%%%
\begin{figure*}[t]
\begin{center}
\includegraphics[width=17cm]{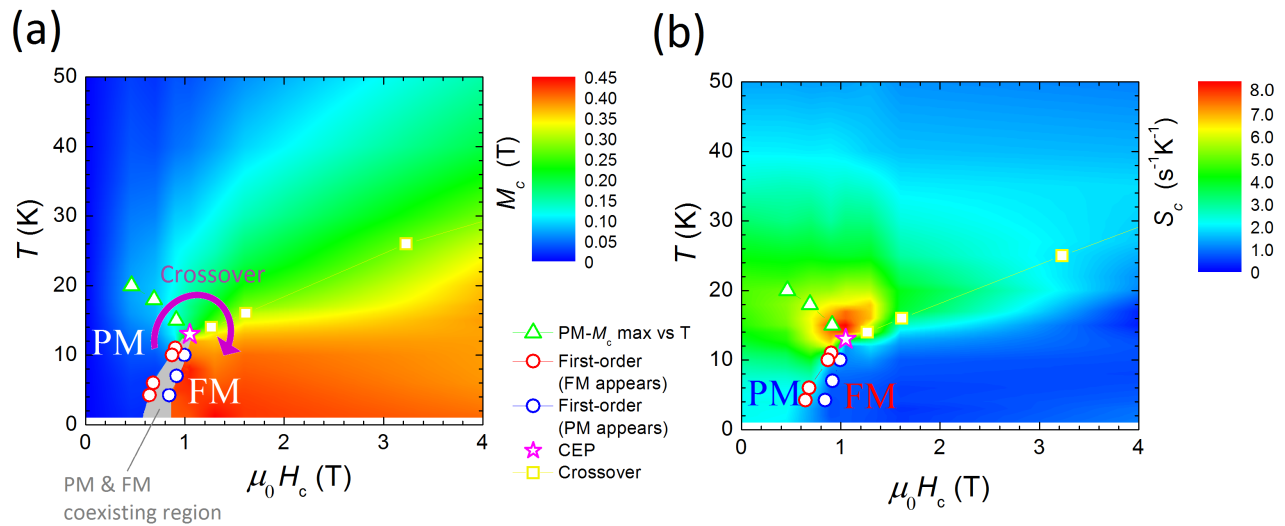}
\end{center}
\caption{(Color online) (a) Magnetization along the $c$-axis $M_c$ in UCoAl is shown by contour plot in the $H_c$ - $T$ phase diagram. In the phase diagram, the red (blue) circle symbol denotes the point where the FM component starts to appear (disappear) by the first-order transition between the PM to FM state. The surrounded area by the red and blue circle symbols (gray colored region) shows the region where the PM and FM components coexist. The light green triangle symbol denotes the point of the temperature $T_{\rm max}$ where $M_c$ has a broad maximum in the PM region. The yellow square symbol denotes the point of crossover determined from the maximum of $\partial M_c/\partial T$. The star symbol denotes the CEP determined as ($\mu_0 H_c$, $T$)$_{\rm CEP} \sim$ (1 T, 12 K). Passing outside the CEP along the purple arrow, $M_c$ changes continuously from the PM to FM states.
(b) Magnetic fluctuation along the $c$-axis $S_c$ in UCoAl is shown by contour plot in the $H_c$ - $T$ phase diagram.
}
\label{fig10}
\end{figure*}
%%%%%%%%%%%%%%%%%%%  FIG10   %%%%%%%%%%%%%%%%%%%%%%%%%%%%%%%%%%%
Figure 9 shows the $T$ dependence of $M_c$ (upper panel) and $S_c$ (lower panel) at fixed several angles $\theta$ (i.e.$H_c = H\cos\theta$). 
In the PM region, $M_c$ shows the broad maximum around 20 K, defined as $T_{\rm max}$. 
$T_{\rm max}$ slightly decreases with increasing $H_c$. 
In a region between $\mu_0H_c \sim 0.7$ and 1.0 T, NMR signals from the PM and FM states were observed as shown in Fig. 6, indicative of the phase separation driven by the first-order transition as observed in the $H_c$ scanned measurements. 
Above $\mu_0H_c \sim 1.0$ T, the first-order transition disappears and changes to crossover as shown in Fig. 7. 
At the PM region, $S_c$ shows almost the same peak structure as $M_c$. 
It should be noted that this peak was observed even at $\mu_0H_c = 0$ T, suggesting that the novel longitudinal magnetic fluctuations are present in zero field. 
This unstable ground state with strong longitudinal FM fluctuations leads UCoAl to the MM transition in a very small external field. 
In fact, Yamada {\it et al.} reported that the MM transition and susceptibility-maximum phenomena are explained with the phenomenological spin-fluctuation model for itinerant-electron metamagnetism\cite{Yamada,Yamada2}. 
With increasing $H_c$, the peak of $S_c$ slightly shifts to lower temperatures and its intensity grows. 
When the $H_c$ exceeds the first-order transition field, the peak of $S_c$ rapidly falls down in the FM region. 
It is also noteworthy that $S_c$ of the PM and FM components possesses almost the same values, although $M_c$ of the PM and FM is different in the phase-separation region as observed in the above $H_c$ scanned measurements. 
These are quite unusual, since $S_c$ of the two states is different in most phase-separation (first-order) phenomena\cite{KoyamaPRB}. 
We suggest that anomalous phase-separation might occur, where the magnetic state is fluctuating between the PM and FM states. Magnetic properties in the coexisting region thus deserves further investigations.  

\subsection{Contour plots of $M_c$ and $S_c$}
Based on the $H_c$ and $T$ scanned measurements of $M_c$ and $S_c$, we developed the contour plots of $M_c$ and $S_c$ in the $H_c -T$ plane, which are shown in Fig. 10 (a) and (b), respectively. 
In the figures, the red (blue) circles show the points where the FM (PM) NMR signal appears (disappears) with increasing $H_c$ or decreasing $T$, and thus the gray-colored region surrounded by these symbols indicates the coexistence of the PM and FM components. 
The light green triangle symbol denotes the point where $M_c$ shows the maximum against $T$ in the PM region. 
The crossover points determined with a maximum of $\partial M_c/\partial T$ are denoted as yellow squares, and the CEP is marked as a star point.
 It is shown that $M_c$ changes continuously from the PM to FM states if the system is varied by following the arrow around the CEP although the transition from the PM and FM states is a first-order transition in small fields. 
Furthermore, $S_c$ diverges significantly at the CEP and gradually decays in crossover region. 
The divergence of $S_c$ was also suggested by the measurement of nuclear spin-spin relaxation rate $1/T_2$ in the $^{59}$Co-NMR\cite{Nohara,comment}.  
These are well-known phenomena observed at a gas-liquid transition. 
In contrast, $M_c$ and $S_c$ show a broad maximum around 20 K in the low-field PM state, which originates from the specific structure of the density of states near the Fermi energy $E_{\rm F}$, and the maximum merges with the CEP with increasing $H_c$. 
The peak structure of $M_c$ and $S_c$ in the PM region is a characteristic feature of itinerant-electron metamagnets, but has not been observed in a gas-liquid transition.

 %%%%%%%%%%%%%%%%%%%%%%%  FIG11   %%%%%%%%%%%%%%%%%%%%%%%%%%%%%%%%%%%
\begin{figure}[t]
\begin{center}
\includegraphics[width=9cm]{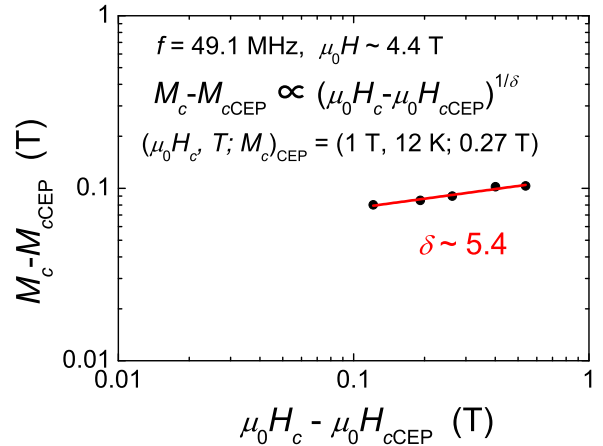}
\end{center}
\caption{(Color online) The power-law fitting of $M_c$ vs $H_c$ on logarithmic scales, where the CEP determined as ($\mu_0~H_c$, $T$; $M_c$)$_{\rm CEP}$ = (1 T, 12 K; 0.27 T). The fitting provided the critical exponent as $\delta$ $\sim$ 5.4. 
}
\label{fig11}
\end{figure}
%%%%%%%%%%%%%%%%%%%%%%%  FIG11   %%%%%%%%%%%%%%%%%%%%%%%%%%%%%%%%%%%

 %%%%%%%%%%%%%%%%%%%%%%%  FIG12   %%%%%%%%%%%%%%%%%%%%%%%%%%%%%%%%%%%
\begin{figure}[t]
\begin{center}
\includegraphics[width=9cm]{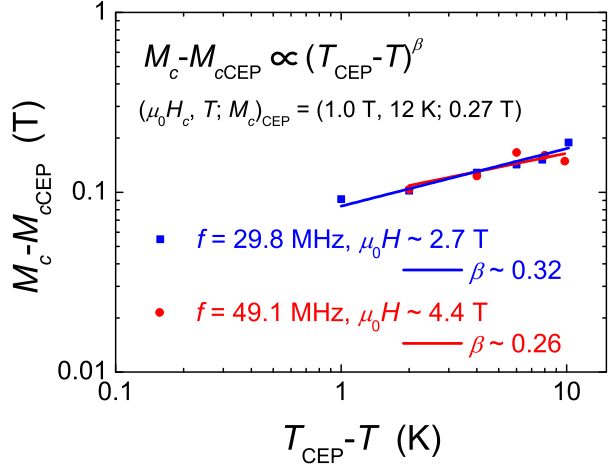}
\end{center}
\caption{(Color online) The power-law fitting of $M_c$ vs $T$ on logarithmic scales, where the CEP determined as ($\mu_0~H_c$, $T$; $M_c$)$_{\rm CEP}$ = (1 T, 12 K; 0.27 T). The fitting provided the critical exponent as $\beta$ $\sim$ 0.26. 
Two cases of $f$ = 29.8 MHz ($\mu_0 H$ $\sim$ 2.7 T) and $f$ = 49.1 MHz ($\mu_0 H$ $\sim$ 4.4 T) provided the almost same fitting result.
}
\label{fig12}
\end{figure}
%%%%%%%%%%%%%%%%%%%%%%%  FIG12   %%%%%%%%%%%%%%%%%%%%%%%%%%%%%%%%%%%

 %%%%%%%%%%%%%%%%%%%%%%%  FIG13   %%%%%%%%%%%%%%%%%%%%%%%%%%%%%%%%%%%
\begin{figure}[t]
\begin{center}
\includegraphics[width=9cm]{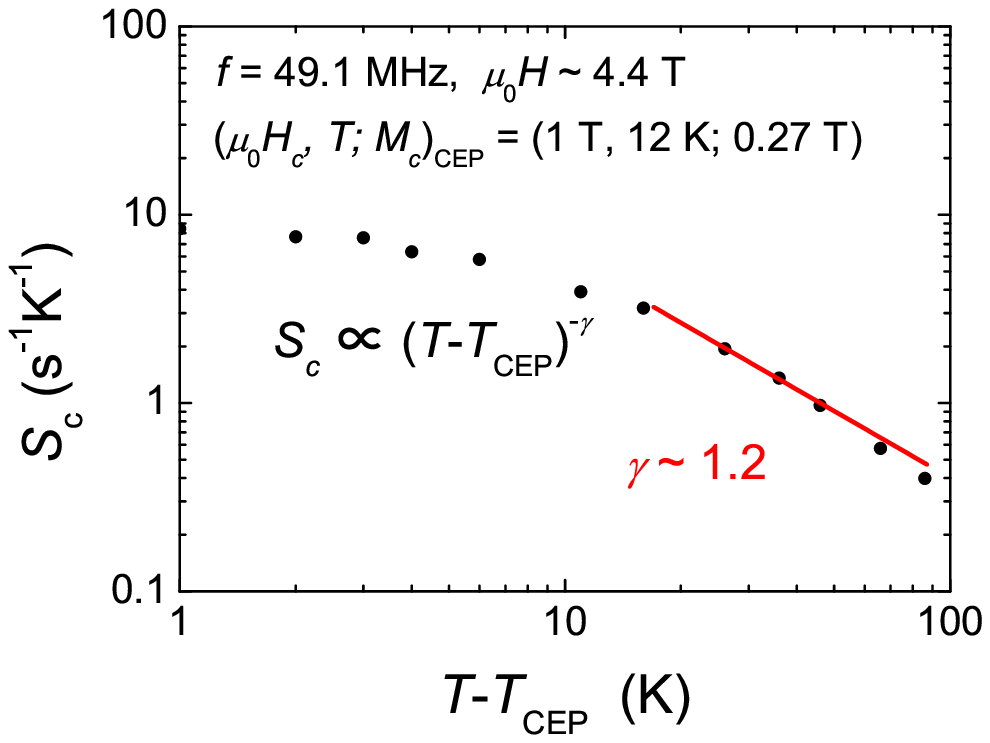}
\end{center}
\caption{(Color online) The power-law fitting of $S_c$ vs $T$ on logarithmic scales, where the CEP determined as ($\mu_0~H_c$, $T$; $M_c$)$_{\rm CEP}$ = (1 T, 12 K; 0.27 T). The fitting provided the critical exponent as $\gamma$ $\sim$ 1.2. 
}
\label{fig13}
\end{figure}
%%%%%%%%%%%%%%%%%%%%%%%  FIG13   %%%%%%%%%%%%%%%%%%%%%%%%%%%%%%%%%%%

 %%%%%%%%%%%%%%%%%%%%%%%  FIG14   %%%%%%%%%%%%%%%%%%%%%%%%%%%%%%%%%%%
\begin{figure}[t]
\begin{center}
\includegraphics[width=8cm]{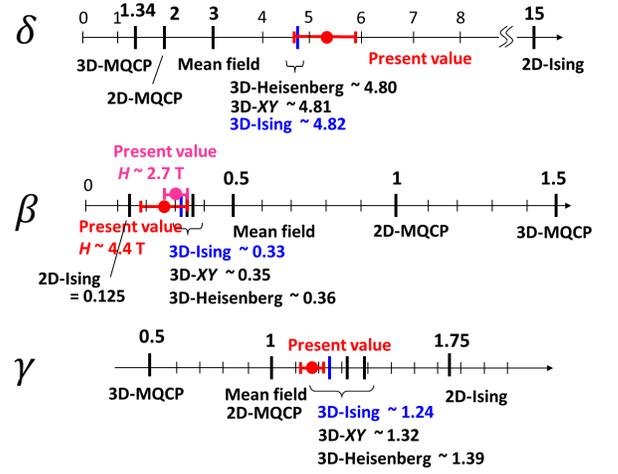}
\end{center}
\caption{(Color online) Comparison of the critical exponents ($\delta$, $\beta$, $\gamma$) of the present case with those of the known universality classes (mean-field, 2D-Ising, 3D-Ising, 3D-$XY$, 3D-Heisenberg, 2D-marginal quantum critical point (2D-MQCP) and 3D-marginal quantum critical point (3D-MQCP) model).
}
\label{fig14}
\end{figure}
%%%%%%%%%%%%%%%%%%%%%%%%  FIG14   %%%%%%%%%%%%%%%%%%%%%%%%%%%%%%%%%%%
\subsection{Critical phenomena around CEP}
In general, the critical phenomena close to the CEP has been analysed on the basis of ``critical exponents".
In the estimation of the critical exponents, we set the CEP as ($\mu_0 H_c$, $T$; $M_c$)$_{\rm CEP}$ = (1 T, 12 K; 0.27 T).
In addition, we used low-energy dynamical susceptibility $S_c$ for the $\gamma$ estimation, except for data points close to the CEP.
This is because the divergence of the susceptibility is sensitively affected by the ambiguities in the determination of $H_c$ and misalignment of the sample, but the value of $\gamma$ would be reliable if a wide temperature range is taken for the estimation. 
The critical exponents were estimated as ($\delta$, $\beta$, $\gamma$) $\sim$ (5.4, 0.26, 1.2) from the fitting shown in Figs. 11, 12 and 13. 
Note that these results almost satisfy the scaling relation $\gamma \sim \beta (\delta-1)$ indicating that the values are evaluated reasonably. The critical exponents ($\delta$, $\beta$, $\gamma$) are plotted along with those of the known universality classes in Fig. 14. 
We found that the universality class observed around the CEP in UCoAl is close to the 3-D critical classes (3D-Ising, 3D-$XY$ and 3D-Heisenberg). However, since UCoAl possesses the strong Ising anisotropy in the static and dynamic magnetic properties, it is reasonable to conclude that UCoAl exhibits a 3D-Ising one, which is the same universality class as a gas-liquid transition. 
The similar universality class was reported in the critical behavior of the conductivity near the 3-D Mott system of Cr-doped V$_2$O$_3$\cite{Limelette}.     
 In contrast, an unconventional critical behavior was reported at the Mott transition occurring in a quasi-2-D organic conductor, probably due to low dimensionality of the system.\cite{Kagawa} 
Therefore, the critical behavior at the finite-temperature MM CEP occurring in UCoAl is a textbook example of the 3D-Ising universality, but the critical behavior when the CEP is tuned to zero temperature [so called quantum critical end point (QCEP)] deserve to be investigated since an unconventional universality featured by the topological transition of Fermi surfaces was suggested at the QCEP\cite{Yamaji}. 

\section{Conclusion}
We derived $c$-axis magnetization $M_c$ and its fluctuation $S_c$ as a function of $H_c$ and $T$ from $^{27}$Al-NMR measurement for single-crystal UCoAl. 
The NMR measurements revealed that UCoAl possesses the 3D FM fluctuations with the strong Ising-type anisotropy. 
Based on the $H_c$ and $T$ scanned measurements, the contour plot of $M_c$ and $S_c$ are developed, and the divergence of $S_c$, which is an anticipated behavior at the CEP, is shown. 
The critical exponents near the CEP of the itinerant MM transition are evaluated, and are found to be categorized to the 3D-Ising universality, which is the same universality class observed in gas-liquid and 3-D Mott transitions.

\begin{acknowledgments}
The authors thank H. Ikeda, H. Kotegawa, and H. Nohara for valuable discussions, and T. Asai, Y. Ihara, Y. Nakai, S. Yonezawa, and Y. Maeno for experimental support and valuable discussions. This work was partially supported by Kyoto Univ. LTM Centre, the ``Heavy Electrons'' Grant-in-Aid for Scientific Research on Innovative Areas  (No. 20102006, No. 21102510, and No. 20102008) from the Ministry of Education, Culture, Sports, Science, and Technology (MEXT) of Japan, a Grant-in-Aid for the Global COE Program ``The Next Generation of Physics, Spun from Universality and Emergence'' from MEXT of Japan, and a Grant-in-Aid for Scientific Research from the Japan Society for Promotion of Science (JSPS).
\end{acknowledgments}

\bibliographystyle{apsrev}
\bibliography{etc,UCoGe}

\end{document}